\documentclass{webofc}
\usepackage[varg]{txfonts}
\usepackage{hyperref}
\usepackage{url}
\hypersetup{colorlinks=true,citecolor=blue,urlcolor=blue,linkcolor=blue}
\usepackage[compact]{titlesec}
\titlespacing*{\section}{0pt}{*1.5}{*0.5}  
\titlespacing*{\subsection}{0pt}{*1.2}{*0.4}

\newcommand{\eq}[1]{\begin{align} #1 \end{align}}
\newcommand{\mean}[1]{\langle #1 \rangle}

\newcommand{\sNN}{\sqrt{s_{\rm NN}}}

\newcommand{\rr}{{\bf r}}

\newcommand{\etacut}{{\eta_{\rm cut}}}

\newcommand{\Nchprim}{N^{\rm prim}_{\rm ch}}

\begin{document}

\title{Probing QGP using local charge fluctuations in heavy-ion collisions}

\author{\firstname{Jonathan} \lastname{Parra}\inst{1}\fnsep\thanks{\email{jparra7@uh.edu}} \and
        \firstname{Roman} \lastname{Poberezhniuk}\inst{1,3}\fnsep\thanks{\email{rpoberez@central.uh.edu}} \and
        \firstname{Volker} \lastname{Koch}\inst{2}
       \and
        \firstname{Claudia} \lastname{Ratti}\inst{1}
        \and
        \firstname{Volodymyr} \lastname{Vovchenko}\inst{1}}

\institute{Department of Physics, University of Houston, Houston, TX 77204, USA
\and
           Nuclear Science Division, Lawrence Berkeley National Laboratory, 1 Cyclotron Road, Berkeley, CA 94720, USA
\and
           Bogolyubov Institute for Theoretical Physics, 03680 Kyiv, Ukraine}

\abstract{\noindent
We revisit the D-measure of event-by-event net-electric charge fluctuations, an idea first introduced over 20 years ago as a potential signature for the presence of quark-gluon plasma (QGP) in heavy-ion collisions. We developed a quantitative framework that incorporates resonance-decay effects, global and local charge conservation, and experimental kinematic cuts. 
Folding these effects into a formalism of density correlations yields an improved expression for the $D$-measure inside acceptance. 
We make comparisons with ALICE data for Pb-Pb collisions at $\sNN=2.76$~TeV. We find that a hadron-gas scenario can describe the data only for a very short charge conservation range, while a QGP scenario is in good agreement with the measurement and is relatively insensitive to the charge conservation range. A Bayesian comparison of both scenarios shows moderate evidence for freeze-out of charge fluctuations in the QGP phase.}

\maketitle

\section{Introduction}
The D-measure of event-by-event net-charge fluctuations is defined as the variance of net electric charge normalized by the mean charge multiplicity:
\eq{\label{eq:D}
D=4\frac{\kappa_2[Q]}{\mean{N_{\rm ch}}},
}
It was proposed as a potential signature for QGP creation in heavy-ion collisions~\cite{JeonKochPRL,AsakawaHeinzPRL}. Due to fractional quark charges, one expects suppressed fluctuations in a deconfined QGP relative to a hadron gas (HG)~(see \cite{Asakawa:2015ybt,Harris:2023tti} for review). 
Measurements were performed at RHIC~\cite{STAR:2008szd} and LHC~\cite{ALICE:2012xnj} but in the absence of quantitative calculations, conclusions remained elusive. 
Grand-canonical ensemble (GCE) estimates for a gas of hadrons yield $D_{HG}\approx 2.8-4$. 
For QGP, a value for $D_{\rm QGP} \approx 1-1.5$ was estimated, three times smaller than the HG value. 

However, the underlying assumptions in the GCE are inaccurate in heavy-ion collisions. Due to the presence of causally disconnected regions of the fireball, there may be charges which are not causally correlated, implying local charge conservation.
Additionally, decays of neutral resonances into charged particles introduces new correlations to the system that have to be accounted for. 
In this work, we develop a formalism that models these effects.

\section{Formalism}
We begin by quantifying fluctuations in the hadronization stage by $\omega=\frac{\kappa_2[Q]}{\mean{\Nchprim}}$,
where $\kappa_2[Q]$ is the net-charge variance and $\mean{\Nchprim}$ is the mean charge multiplicity \emph{before} resonance decays. The mean multiplicity is in chemical equilibrium and is determined by the Hadron Resonance Gas (HRG) model. The net-charge variance is a conserved quantity and can only change through diffusion in a finite acceptance. This process may be slow enough that equilibrium is not maintained and the net-charge variance may decouple before the mean charge multiplicities do, and thus, may freeze out in the hadron gas phase or as early as in the QGP phase. Particles detected later on may contain the memory of where this occurs. 

For a hadron gas scenario, the Maxwell-Boltzmann limit (Poisson statistics) gives $\omega_{\rm HG} \approx 1$, but when considering Bose-Einstein statistics for pions and the presence of multi-charged hadrons, an HRG calculation gives $\omega_{\rm HG} \approx 1.1$. For a QGP scenario, the difficulty lies in estimating the mean multiplicity of charged hadrons $\mean{\Nchprim}$.
Thus, we utilize the approximately isentropic evolution between the freeze-out of charge fluctuations and hadronization,
and the fact that in the grand-canonical ensemble,  $\kappa_2[Q] = V \chi_2^Q$. Using this, we can divide and multiply $\omega$ by entropy $S$. Hence,
\eq{\label{eq:omega}
\omega 
& = \frac{\kappa_2[Q]}{\mean{N_{\rm ch}^{\rm prim}}} 
= \frac{V \chi_2^Q}{S} \frac{S}{\mean{N_{\rm ch}^{\rm prim}}}  = \frac{\chi_2^Q}{s} \frac{S}{\mean{N_{\rm ch}}} \frac{\mean{N_{\rm ch}}}{\mean{N_{\rm ch}^{\rm prim}}}.
}
Charge susceptibility $\chi_2^Q$ and entropy density $s$ are given in the QGP by the Stefan-Bolzmann limit of massless quarks: $(\chi_2^{Q})_{\rm QGP} / T^3 = 2/3$ and $s_{\rm QGP}/ T^3 = 19\pi^2 / 9$.  
For entropy per hadron, we use a recent analysis which gives $S/\mean{N_{ch}} = 6.7 \pm 0.8$~\cite{Hanus:2019fnc}. Finally, $\mean{N_{\rm ch}}/\mean{N_{\rm ch}^{\rm prim}}$, which we call $
\gamma_Q$, is the ratio of final and primordial charge multiplicities.
A thermal model estimate, which we employ, gives $\gamma_Q = 1.67$. 
Altogether one obtains $\omega_{\rm QGP} = 0.36 \pm 0.04$,
where the uncertainty comes solely from $S/\mean{N_{\rm ch}}$. 
These estimates for $\omega$ for the HG and QGP scenarios are further justified by a lattice QCD based calculation of the temperature dependence of $\omega$~(see Fig. 1 in \cite{Parra2025} for details).

Next, we construct a charge susceptibility $\chi_2^Q$ in a form that encapsulates correlations among charges, such as resonance decays. 
First, we write the susceptibility as $\chi_2^{Q} = \omega \mean{n_{\rm ch}^{\rm prim}}$, where $\mean{n_{\rm ch}^{\rm prim}}$ is the charged particle density. 
We then decompose $\chi_2^Q$ into two terms which correspond to self-correlations and 2-particle correlations, respectively: \eq{\label{eq:}
\chi_2^{Q} = \mean{n_{\rm ch}^{\rm prim}} + \varphi_2^{Q, \rm prim}.} Before any decays occur, the strength of correlations is parameterized by $\omega$. Hence, $\chi_2^{Q} = \omega \mean{n_{\rm ch}^{\rm prim}}$ and 
$\varphi_2^{Q, \rm prim} = (\omega - 1) \mean{n_{\rm ch}^{\rm prim}}$. After resonance decays, the net charge (and thus $\chi_2^Q$)  remain conserved, but the multiplicities increase by a factor of $\gamma_Q$ ($\approx 1.67$). Thus, $\mean{n_{\rm ch}} = \gamma_Q \mean{n_{\rm ch}^{\rm prim}}$. This introduces additional correlations which induce a rebalancing of self- and 2-particle correlations. We write the final charged susceptibility as $
\chi_2^Q = \mean{n_{\rm ch}} + \varphi_2^{Q}$
where, \eq{\label{eq:varphifinal}
\varphi_2^{Q} = \left( \frac{\omega}{\gamma_Q} - 1 \right) \mean{n_{\rm ch}}.
}

To model exact charge conservation, we use 
the 2-point correlator $\mathcal{C}^Q_2(\rr_1,\rr_2)$ which was recently derived for a thermal system with local charge conservation in spatial rapidity coordinates, $(\eta_1,\eta_2)$. It consists of local and balancing terms~\cite{Vovchenko:2024pvk}:
\eq{
\mathcal{C}^Q_2(\eta_1,\eta_2) = \chi_2^Q \left[ \delta(\eta_1 - \eta_2) - \frac{\varkappa(\eta_1,\eta_2)}{V_{\rm tot}} \right].
}
Here $\chi_2^Q$ is the grand-canonical charge susceptibility, $V_{\rm tot} = \int d \rr$ is the total system volume, and $\varkappa(\rr_1,\rr_2)$ 
is the local charge conservation kernel which is modeled as a Gaussian function. The Gaussian width, $\sigma$, is related to an effective conservation volume $V_c/V_{tot}$.
Integrating the 2-point correlator over a subvolume yields the net-charge variance of particles emitted from that subvolume.
To account for collective flow and kinematical cuts in momentum space, we employ the binomial model for acceptance probability, which is the probability $p(\eta)$ that a particle with some spatial rapidity  $\eta$ will end up in momentum acceptance. We then employ the blast-wave model fitted to single-particle spectra~\cite{ALICE:2013mez} to calculate the acceptance probabilities.

Combining the aforementioned effects yields an expression for the D-measure inside momentum acceptance:
\eq{\label{eq:Dformalism}
D=4\frac{\kappa_2[Q]}{\mean{N_{\rm ch}}}=4 \left\{ 1 - \left( 1- \frac{\omega}{\gamma_Q} \right) \frac{\mean{p^2(\eta)}}{\mean{p(\eta)}} - \frac{\omega}{\gamma_Q} \frac{\mean{p(\eta_1) p(\eta_2)}_\varkappa}{\mean{p(\eta)}} \right\},
}
which contains terms that correspond to the Skellam baseline, short-range two-particle correlations (due to decays or hadronization of QGP), and long-range correlations.

\begin{figure}[b!]
  \centering
  \sidecaption
  \includegraphics[width=.55\textwidth,clip]{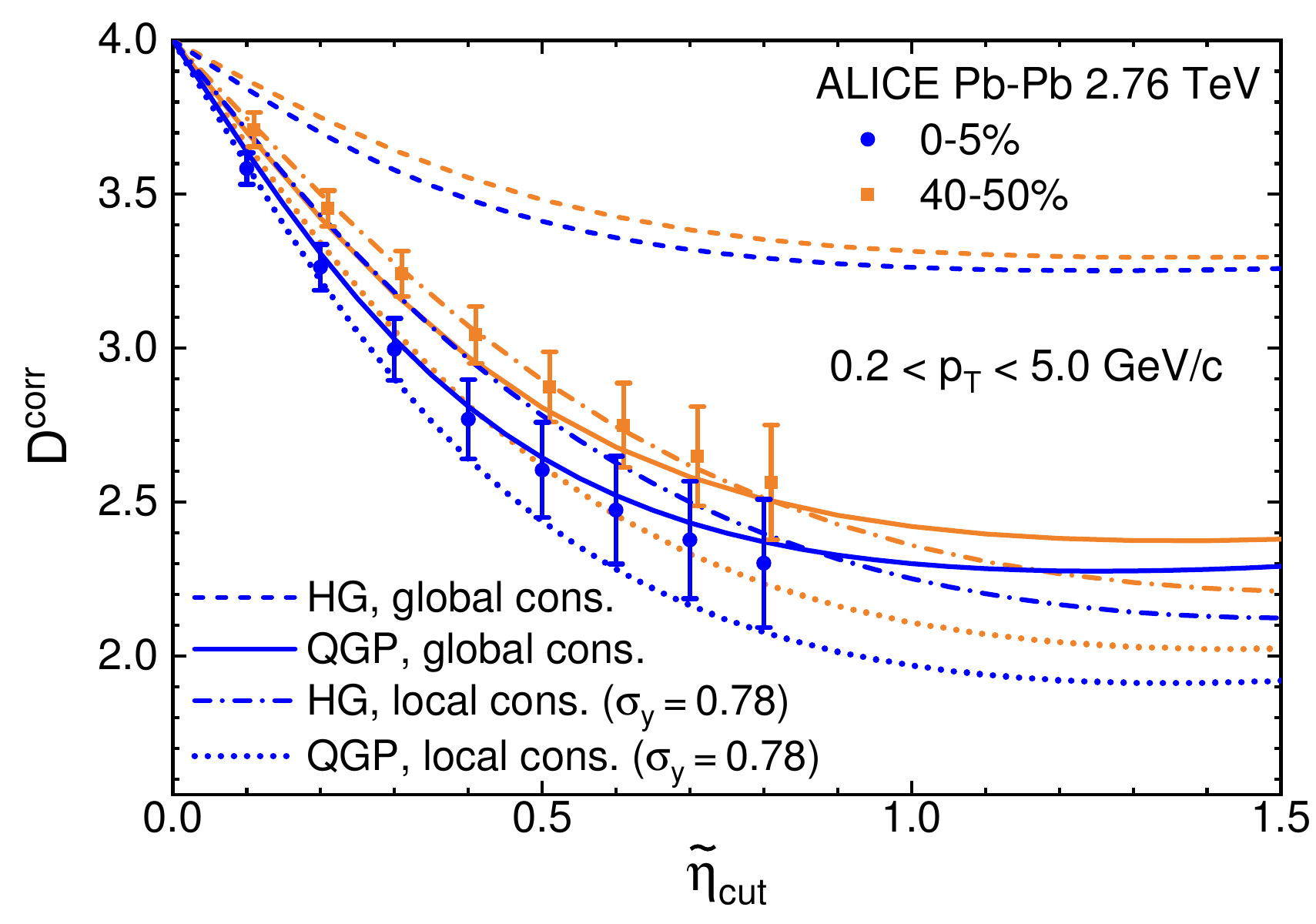}
  \caption{Corrected D-measure as a function of the pseudorapidity cut $\tilde{\eta}_{\rm cut}$ in 0-5\% (blue curves) and 40-50\% (orange curves) collisions. The QGP and HG scenarios under global charge conservation are represented by solid and dashed lines, respectively, while those under local charge conservation are shown as dotted and dash-dotted lines. Symbols show the experimental data from ALICE~\cite{ALICE:2012xnj}.}
  \label{fig:alice}
\end{figure}
\section{Comparison to ALICE data at $\sNN=2.76$~TeV}
The ALICE Collaboration published data for the D-measure in Pb-Pb collisions at $\sNN=2.76$~TeV \cite{ALICE:2012xnj}. In Fig.~\ref{fig:alice} we plot the curves obtained from Eq.~\eqref{eq:Dformalism} as functions of pseudorapidity cut $\tilde \eta_{\rm cut}$ for the HG ($\omega\approx1.1$) and QGP ($\omega\approx0.36$) scenarios. 
The results are presented for 0-5\% and 40-50\% centralities, using the appropriate blast-wave parameters.
To elucidate the effect of local charge conservation, we performed the calculation for global charge conservation~($V_c = V_{\rm tot}$), and a representative value of local charge conservation, $V_c = 0.2 V_{\rm tot}$~($\sigma_y = 0.78)$, motivated by Ref.~\cite{Vovchenko:2024pvk}. 
The HG scenario describes the data only for a short range of local charge conservation  (dash-dotted blue curve), while the QGP scenario is in excellent agreement with the data (solid and dotted blue curves) at any range. 
This is attributed to the dependence of the local conservation term on $\omega$, which for the QGP is much smaller than for the HG.
The calculations reproduce centrality dependence trend in the data.

\section{Bayesian analysis}
\begin{figure}[h!]
  \centering
  \sidecaption
  \includegraphics[width=7cm,clip]{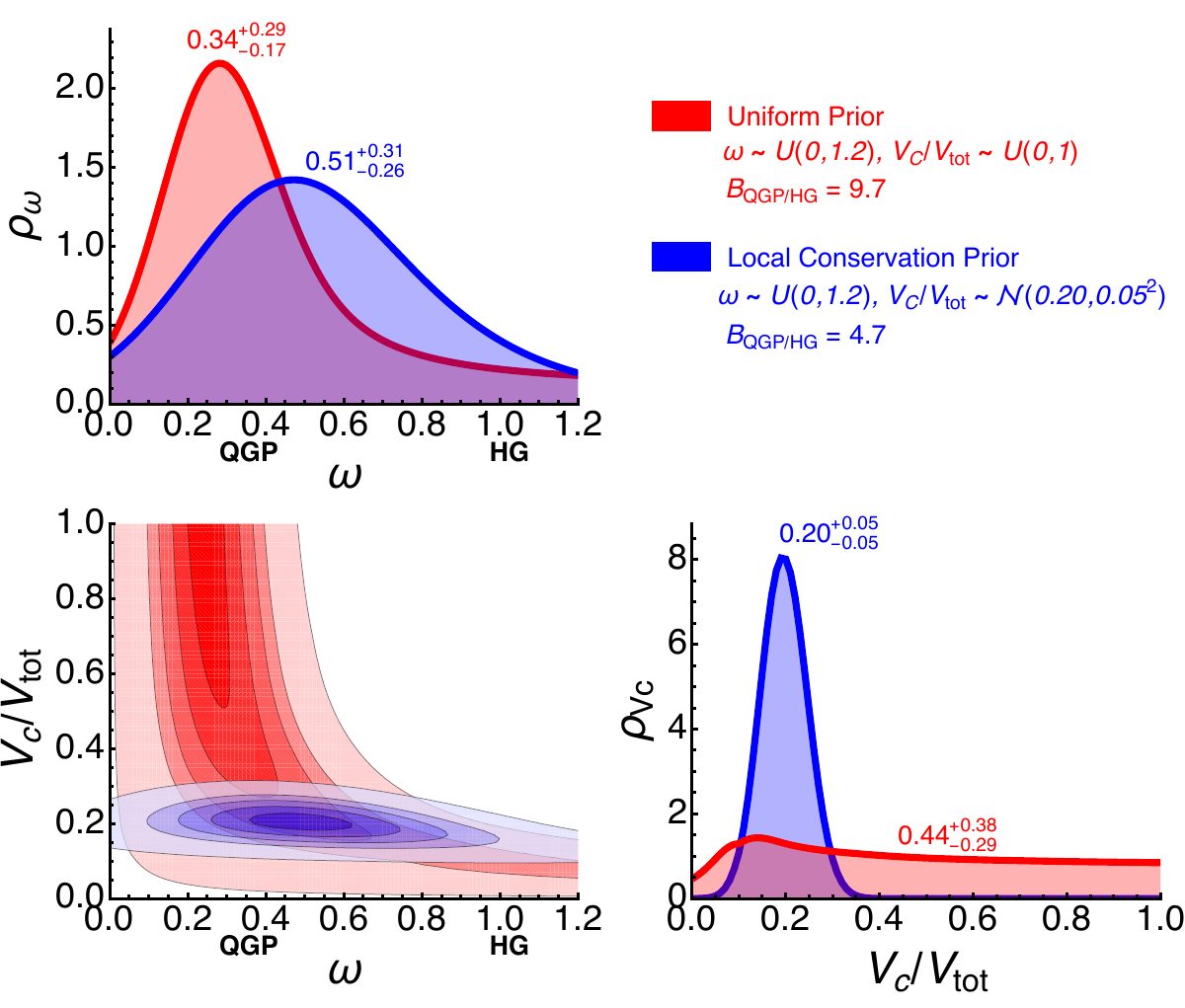}
  \caption{Posterior distributions for $\omega$ and $V_c/V_{\rm tot}$ under uniform (red) and local-conservation (blue) priors for $V_c/V_{\rm tot}$. 
  The Bayes factor compares the QGP and HG charge fluctuations freeze-out scenarios, yielding moderate evidence for QGP.}
  \label{fig:corner}
\end{figure}

To further investigate which scenario fits the data best, we perform a two-parameter Bayesian analysis in $(\omega, V_c/V_{\rm tot})$ using the smallest and largest $\etacut$ data points for 0-5\% centrality. 
Here, we vary $\omega \in [0,1.2]$, which covers all possible intermediate regimes between HG and QGP, and even vanishing local fluctuations in coordinate space. 
We take uniform priors $\omega\!\sim\!U(0,1.2)$ and $V_c/V_{\rm tot}\!\sim\!U(0,1)$ which yield a posterior that moderately prefers QGP over HG with Bayes factor $B_{\rm QGP/HG}\!\approx\!9.7$. 
A prior preferring local conservation, $V_c/V_{\rm tot}\!\sim\!\mathcal{N}(0.20,0.05^2)$, still favors QGP with $B\!\approx\!4.7$. 
These Bayes factor values are indicative of \emph{moderate} evidence~\cite{jeffreys1961theory} for the freeze-out of charge fluctuations in the QGP over the HG scenario, see Fig.~\ref{fig:corner} for the full result.

\section{Conclusions and Outlook}
We developed a new formalism that unifies suppression of net-charge fluctuations in the QGP phase, resonance decay effects, local charge conservation, and kinematic acceptance in a single quantitative framework. We compared our results to ALICE Run 1 data and found that a HG scenario describes the data only when taking a very local range of charge conservation into account, while a QGP scenario is in good agreement at any range. A Bayesian analysis of both scenarios reveals moderate evidence for freeze-out of fluctuations in the QGP phase. We look forward to the release of LHC Run 2 data to further test our framework.
\section*{Acknowledgments}
We thank Mesut Arslandok for fruitful discussions. This work was supported by the U.S. Department of Energy, Office of Science, Office of Nuclear Physics, under contract numbers DE-SC0022023 (J.P., C.R., V.V.) and DE-AC02-05CH11231 (V.K.). This material is based upon work supported by the National Science Foundation under grants No. PHY-2514763, No. PHY-2208724, and PHY-2116686, and within the framework of the MUSES collaboration, under Grant No. OAC-2103680.

\end{document}